\documentclass[12pt]{article}
\usepackage{amsfonts}
\font\oneeight=cmr10 at 18pt
\font\onesix=cmr10 at 16pt
\font\onefour=cmr10 at 14pt

\newcommand{\vTm}{\vphantom{\mbox{\oneeight I}}}
\newcommand{\vTmm}{\vphantom{\mbox{\onesix I}}}
\newcommand{\vTs}{\vphantom{\mbox{\onefour I}}}

\begin{document}

\baselineskip=20pt

\newfont{\elevenmib}{cmmib10 scaled\magstep1}
\newcommand{\preprint}{
    \begin{flushleft}
      \elevenmib Yukawa\, Institute\, Kyoto\\
    \end{flushleft}\vspace{-1.3cm}
    \begin{flushright}\normalsize  \sf
      YITP-03-09\\
      {\tt hep-th/0305120} \\ May 2003
    \end{flushright}}
\newcommand{\Title}[1]{{\baselineskip=26pt
    \begin{center} \Large \bf #1 \\ \ \\ \end{center}}}
\newcommand{\Author}{\begin{center}
    \large \bf O.~Ragnisco${}^a$ and R.~Sasaki${}^b$ \end{center}}
\newcommand{\Address}{\begin{center}
      $^a$ Department of Physics,  Roma Tre University,\\
      Via Vasca Navale 84, Rome I-00146,  Italy\\
      ${}^b$ Yukawa Institute for Theoretical Physics,\\
      Kyoto University, Kyoto 606-8502, Japan
    \end{center}}
\newcommand{\Accepted}[1]{\begin{center}
    {\large \sf #1}\\ \vspace{1mm}{\small \sf Accepted for Publication}
    \end{center}}

\preprint
\thispagestyle{empty}
\bigskip\bigskip\bigskip

\Title{Quantum vs Classical  Integrability\\ in Ruijsenaars-Schneider
Systems}
\Author

\Address
\vspace{1cm}

\begin{abstract}
The relationship (resemblance and/or contrast) between
quantum and classical integrability
in Ruijsenaars-Schneider systems, which are  one parameter deformation of
Calogero-Moser systems, is addressed.
Many remarkable properties of
classical Calogero and Sutherland systems (based on {\em any\/} root system)
at equilibrium are reported in a previous paper (Corrigan-Sasaki).
For example, the minimum energies, frequencies of small oscillations and
the eigenvalues of Lax pair matrices at equilibrium are all ``integer valued".
In this paper we report that similar
features and results hold for the Ruijsenaars-Schneider
type of integrable systems based on the {\em classical\/} root systems.
\end{abstract}

\section{Introduction}
\label{intro}
\setcounter{equation}{0}

If a many-body dynamical system is (Liouville) integrable at both classical
and quantum levels, the classical system appears to share many `quantum'
features. For example, the frequencies of small oscillations near equilibrium
are `quantised' together with the eigenvalues of the associated Lax matrices
at the equilibrium. This phenomenon of close relation/contrast between quantum
and classical integrability has been explored extensively by
Corrigan-Sasaki \cite{cs} (this paper will be referred to as I hereafter)
for Calogero-Moser (C-M) systems \cite{Cal,Sut, CalMo} based on any root
system
\cite{OP1,DHoker_Phong,bms}.
  In this paper we show that similar phenomenon occurs
  in Ruijsenaars-Schneider \cite{Ruij-Sch,vanDiejen1,vanDiejen2,vanDiejen0}
systems, which are  one parameter {\em deformation\/}
  of C-M 
systems. The equations determining equilibrium can be presented in a form
similar to {\em Bethe ansatz\/} equations.
The equilibrium positions are described as the zeros of certain
`deformed' classical polynomials, which tend to the Hermite, Laguerre and
Jacobi polynomials in the C-M limit \cite{cs}.
The frequencies of small oscillations at
equilibrium are either integers or a sort of `{\em deformed\/}' integers
according to the potential is rational or trigonometric, respectively. In the
C-M limits, these frequencies tend to those presented in I, which
provide non-trivial support for the current results.

This paper is organised as follows. In section 2, the Ruijsenaars-Schneider
(R-S)
systems  \cite{Ruij-Sch} are briefly introduced and the formulas for
evaluating the frequencies of small oscillations at equilibrium are derived.
In most cases, the evaluation of the frequencies is done numerically.
In section 3, the rational Ruijsenaars-Schneider systems with  two
types of confining potentials for the classical root systems $A$, $B$, $C$,
$BC$ and
$D$
\cite{vanDiejen1,vanDiejen2} are recapitulated. The formulas of the
frequencies of the small oscillations are presented and compared with the
corresponding results in Calogero systems \cite{Cal} given in I. In section 4
the trigonometric R-S systems for the classical root systems
are reviewed briefly. The formulas of the frequencies of small oscillations
at equilibrium are presented and compared with the results of the Sutherland
systems \cite{Sut} given in I.

\section{Ruijsenaars-type systems}
\label{R-typel}
\setcounter{equation}{0}
The Ruijsenaars-Schneider  systems are
``discrete" versions of the C-M 
systems
\cite{Cal,Sut,CalMo}, that is the momentum variables appear in the Hamiltonian
not as polynomials but as exponential (hyperbolic) functions.
In quantum theoretical setting this would mean that the operator
(an ``analytic difference operator'', according to S.\, Ruijsenaars)
$e^{\pm\beta p}=e^{\mp i\beta\hbar {\partial/\partial q}}$
causes a shift of the wavefunction by
a finite unit $\beta\hbar$  in the imaginary direction, {\em i.e.\/}
$\psi(q)$ to $\psi(q\mp i\beta\hbar)$.
The parameter $\beta$ has the dimensions of a momentum$\!{}^{-1}$
and can be expressed as
$\beta=1/mc$, in which $m$ and $c$ are constants of the dimensions of the
{\em mass\/} and the {\em velocity\/}, respectively.
The R-S systems can also be considered as  a one-parameter ($c$)
deformation of
the C-M systems (which correspond to the $c\to\infty$ limit)
and are sometimes referred to, somehow misleadingly,
as ``relativistic" Calogero-Moser
systems. See \cite{Braden-Sasaki} for comments on this point.

The dynamical variables of the classical Ruijsenaars-Schneider
model are the coordinates
\(\{q_{j}|\,j=1,\ldots,r\}\) and their canonically conjugate momenta
\(\{p_{j}|\,j=1,\ldots,r\}\), with the Poisson bracket relations:
\begin{eqnarray*}
  \{q_{j},p_{k}\}=\delta_{j\,k},\qquad \{q_{j},q_{k}\}=
    \{p_{j},p_{k}\}=0.
\end{eqnarray*}
These will be  denoted by vectors in \(\mathbb{R}^{r}\)
\[
    q=(q_{1},\ldots,q_{r}),\qquad p=(p_{1},\ldots,p_{r}),
\]
in which $r$ is the number of particles and it is also the
rank of the underlying root system $\Delta$.
In this paper we will discuss those models associated
with the {\em classical\/}
root systems, namely the $A$, $B$, $C$, $D$ and $BC$.
The fact that all the roots are neatly expressed in terms of the
orthonormal basis of $\mathbb{R}^r$ makes formulation
much simpler than those systems based on exceptional root systems.
Throughout this paper we adopt the convention of
$\beta\equiv1$ for convenience.

Following Ruijsenaars-Schneider \cite{Ruij-Sch} and van Diejen
\cite{vanDiejen1}, let us start with the following Hamiltonian
\begin{equation}
H(p,q)=\sum_{j=1}^r\left[ 2\cosh
p_j\,\sqrt{V_j(q)\,{V}_j^*(q)}-(V_j(q)+{V}_j^*(q))\right],
\label{Ham}
\end{equation}
in which ${V}_j^*$ is the complex conjugate of $V_j$.
The form of the function $V_j=V_j(q)$ is determined by
the root system $\Delta$ as:
\begin{eqnarray}
  A: && V_j(q)=w(q_j)\prod_{k\neq j}v(q_j-q_k),
\qquad\qquad j=1,\ldots,r+1,
\label{AVform}\\
B,\ C,\ BC, \ \&\, D: && V_j(q)=w(q_j)\prod_{k\neq j}v(q_j-q_k)v(q_j+q_k),
\quad j=1,\ldots,r.
\label{otherVform}
\end{eqnarray}
The elementary potential functions $v$ and $w$ depend on
the nature of interactions
(rational, trigonometric, etc)  and the root system $\Delta$.

\subsection{Equilibrium position and frequencies of small oscillations}
It is easy to see that the system has a stationary solution
\begin{equation}
p=0,\qquad q=\bar{q},
\label{stasol}
\end{equation}
of the canonical
equations of motion:
\begin{eqnarray}
\dot{q}_j&=&{\partial H(p,q)\over{\partial p_j}}=\sum_{j=1}^r2\sin
p_j\sqrt{V_jV_j^*},\qquad
\dot{p}_j=-{\partial H(p,q)\over{\partial q_j}},
\end{eqnarray}
in which $\bar{q}$ satisfies
\begin{equation}
\left.{\partial H(0,q)\over{\partial q_j}}\right|_{\bar{q}}=0,\qquad
j=1,\ldots,r.
\end{equation}

By expanding the Hamiltonian around the stationary solution (\ref{stasol}),
we obtain
\begin{equation}
H(p,q)=K(p)+P(q)+\mbox{higher order terms in } p,
\end{equation}
in which the kinetic part $K$ is quadratic in $p$
\begin{equation}
K(p)=\sum_{j=1}^r(p_j)^2 a_j,\quad a_j\equiv |V_j(\bar{q})|,
\end{equation}
and the `potential' $P$ is given by
\begin{equation}
P(q)=\sum_{j=1}^r\left(2\sqrt{V_j(q)\,{V}_j^*(q)}-(V_j(q)+{V}_j^*(q))\right)
=-\sum_{j=1}^r\left(\sqrt{V_j(q)}-\sqrt{V_j^*(q)}\right)^2.
\label{potP}
\end{equation}
This should be compared with the  classical potential in
C-M systems,
$V_C=\sum_{j=1}^r(\partial W/\partial q_j)^2$, in which $W$ is the
  {\em prepotential\/}
\cite{bms}.
It is obvious that the equilibrium is achieved at the point(s) in which all the
functions
$V_j$ become {\em real\/} and {\em positive\/}:
\begin{equation}
V_j(\bar{q})=V_j(\bar{q})^*>0,\qquad j=1,2,\ldots,r.
\label{equeq}
\end{equation}
Therefore the ``minimal energy" $P(\bar{q})$ is always 0 in contrast
to the C-M cases.

Let us define the Hessian of the `potential' $P$ at equilibrium as $B_{jk}$:
\begin{equation}
B_{jk}\equiv \left.{\partial^2 P(q)\over
{\partial q_j\partial q_k}}\right|_{\bar{q}},
\quad j,k=1,\ldots,r.
\end{equation}
It is easy to verify that
\begin{equation}
B_{jk}={1\over2}\sum_{l=1}^r
\left.{1\over{a_l}}{\partial IV_l\over{\partial q_j}}
\right|_{\bar{q}}
\left.{\partial IV_l\over{\partial q_k}}
\right|_{\bar{q}},\quad IV_l(q)\equiv -i(V_l(q)-V_l^*(q)),
\quad j,k=1,\ldots,r.
\end{equation}
Thus the small oscillations near the stationary point
  (\ref{stasol}) are described by
the effective quadratic Hamiltonian in $p$ and $q-\bar{q}$:
\begin{equation}
H_{eff}(p,q)=K(p)+{1\over2}\sum_{j,k=1}^rB_{jk}(q-\bar{q})_j(q-\bar{q})_k.
\end{equation}
In terms of a canonical transformation
\begin{equation}
p'_j=\sqrt{2a_j}\,p_j,\quad q'_j=
{1\over\sqrt{2a_j}}(q-\bar{q})_j,\quad j=1,\ldots,r,
\end{equation}
the quadratic Hamiltonian reads
\begin{eqnarray}
H_{eff}(p',q')&=&{1\over2}\sum_{j=1}^r(p'_j)^2+
{1\over2}\sum_{j,k=1}^rB'_{jk}q'_jq'_k,\\
B'_{jk}&\equiv&(W^2)_{jk},
\end{eqnarray}
in which a real {\em symmetric%
\footnote{This property stems from
the structure of the functions  $V_j$, (\ref{AVform}), (\ref{otherVform})
and from the even nature of $v'(x)$.
\/}} $r\times r$ matrix $W$ is defined by
\begin{equation}
W_{jk}\equiv\sqrt{a_j}\,\left.
{\partial IV_k\over{\partial q_j}}\right|_{\bar{q}}{1\over\sqrt{a_k}}.
\end{equation}
The frequencies of small oscillations at the equilibrium are given
simply by the eigenvalues of a matrix $\widetilde{W}$:
\begin{equation}
\widetilde{W}_{jk}\equiv\left.
{\partial IV_k\over{\partial q_j}}\right|_{\bar{q}},
\end{equation}
which are relatively easy to evaluate.

\section{Ruijsenaars-Calogero systems}
\label{R-Cal}
\setcounter{equation}{0}

Here we will discuss the discrete analogue
of the Calogero systems \cite{Cal}, to be called
Ruijsenaars-Calogero systems, which were introduced by
van Diejen for classical root systems only \cite{vanDiejen1,vanDiejen2}.
(For the definition of C-M systems based on any root system and
the associated Lax representation, etc, see \cite{OP1,DHoker_Phong,bms}.)
The original Calogero systems \cite{Cal} have the rational $(1/q^2)$
potential plus the harmonic $(q^2)$  confining potential,
having two coupling constants $g$ and $\omega$ for the systems based on the
simply-laced root systems,
$A$ and $D$, and three couplings $\omega$ and
$g_L$ for the long roots and $g_S$
for the short roots in $B\, (C)$ root system.
(For the rational Calogero systems,
those based on $B$ and $C$ root systems are equivalent.)

The discrete Calogero systems have two varieties (deformations),
according to the number of independent coupling constants.
The first has two (three for the non-simply-laced root systems)
coupling constants $g$ ($g_L$ and $g_S$) and $a$
which corresponds to $\omega$
in the Calogero systems.
This can be called a ``minimal" discretisation of the Calogero systems.
The second is introduced by van Diejen  \cite{vanDiejen1,vanDiejen2} having
three (four for the non-simply-laced root systems) coupling constants $g$
($g_L$ and $g_S$) and $a,\,b$ both of which correspond to $\omega$.
In this case the $B$ and $C$ systems are not equivalent in contrast to
the Calogero systems.
The integrability (classical and quantum) of the latter was discussed
by van Diejen in some detail \cite{vanDiejen1,vanDiejen2}.
Whereas, the former (the minimal discretisation)
is new and its integrability has not been discussed
to the best of our knowledge.
As we will show in the next subsection,  the very orderly spectrum of the
small oscillations would give strong evidence for its integrability.

\subsection{Linear confining potential case}
\label{Rlin}
Let us first write down the explicit forms of the elementary potential
functions
$v$ and $w$.
For the simply-laced root systems $A$ and $D$ the
elementary potential functions
are:
\begin{eqnarray}
\hspace*{-3.7cm}
A,\ D:\quad\ v(x)=1-ig/x,\quad w(x)=a+ix,
\label{adminC}
\end{eqnarray}
in which $a$ and $g$ are real coupling constants assumed to be positive.
For the non-simply-laced root systems $B$ ($C$), $BC$,
they  are:
\begin{eqnarray}
\ B,\ (C):&& v(x)=1-ig_L/x,\quad w(x)=(a+ix)(1-ig_S/x),\label{bminC}\\
\ BC:&& v(x)=1-ig_M/x,
\quad w(x)=(a+ix)(1-ig_S/x)(1-ig_L/2x),
\label{bcminC}
\end{eqnarray}
in which  $g_L$, $g_M$ and $g_S$ are the independent positive coupling
constants for the long, middle and short roots, respectively.  As in the
Calogero case, those based on $B$ and $C$ systems are equivalent. In all
these cases the `potential'
$P$ (\ref{potP}) grows linearly in
$|q|$ as
$|q|\to\infty$.
Except for the $BC$ case, there are simple identities:
\begin{equation}
\sum_{j}\left\{V_j(q)+V_j(q)^*\right\}=const.
\label{Viden}
\end{equation}
Thus the Hamiltonian (\ref{Ham}) can be replaced by a simpler one
\begin{equation}
H'(p,q)=2\sum_{j=1}^r\cosh p_j\,\sqrt{V_j(q)\,{V}_j^*(q)},
\label{Ham2}
\end{equation}
which is usually used as a starting point for the trigonometric (hyperbolic)
interaction theory, see section \ref{R-Sut}.
To be more precise, the identities are:
\begin{equation}
A_r:\qquad \sum_{j=1}^{r+1}\left\{V_j(q)+V_j(q)^*\right\}=2(r+1)a+r(r+1)g,
\end{equation}
and
\begin{equation}
B_r\ (C_r)\ \&\ D_r: \qquad \sum_{j=1}^{r}\left\{V_j(q)+V_j(q)^*\right\}=
2r\left(\vTs a+(r-1)g_L+g_S\right).
\end{equation}
(For $D_r$, $g_L\to g$ and $g_S\equiv0$.)

It is interesting to note that the equations determining the equilibrium
(\ref{equeq}), in general,
can be cast in a form which looks similar to the {\em Bethe
ansatz\/} equation. For example, for the elementary potential
(\ref{adminC})--(\ref{bminC}), (\ref{equeq}) read
\begin{eqnarray}
A_r:\hspace{27mm}\prod_{k\neq
j}{\bar{q}_j-\bar{q}_k-ig\over{\bar{q}_j-\bar{q}_k+ig}}&=&
{a-i\bar{q}_j\over{a+i\bar{q}_j}},\hspace{21mm} j=1,\ldots,r+1,
\label{aminCeq}\\
D_r:\quad\ \prod_{k\neq
j}{\bar{q}_j-\bar{q}_k-ig\over{\bar{q}_j-\bar{q}_k+ig}}
{\bar{q}_j+\bar{q}_k-ig\over{\bar{q}_j+\bar{q}_k+ig}}&=&
{a-i\bar{q}_j\over{a+i\bar{q}_j}},\hspace{21mm} j=1,\ldots,r,
\label{dminCeq}\\
B_r:\ \prod_{k\neq
j}{\bar{q}_j-\bar{q}_k-ig_L\over{\bar{q}_j-\bar{q}_k+ig_L}}
{\bar{q}_j+\bar{q}_k-ig_L\over{\bar{q}_j+\bar{q}_k+ig_L}}&=&
{a-i\bar{q}_j\over{a+i\bar{q}_j}}{\bar{q}_j+ig_S\over{\bar{q}_j-ig_S}},\quad\
j=1,\ldots,r.
\label{bminCeq}
\end{eqnarray}
They determine the zeros of deformed Hermite and Laguerre polynomials.
  For $A_r$, let us
define
\begin{equation}
\bar{q}_j=\sqrt{ag}\,y_j,\quad \delta=g/a,
\end{equation}
and introduce a degree $r+1$ polynomial
\begin{equation}
H_{r+1}^{(\delta)}(x)=2^{r+1}\prod_{j=1}^{r+1}(x-y_j),
\end{equation}
which is a deformation of the Hermite polynomial $H_{r+1}(x)$. For lower $n$,
$H_n^{(\delta)}(x)$ are:
\begin{eqnarray}
H_n^{(\delta)}(x)&=&H_{n}(x),\quad n=0,1,2,\quad\
H_3^{(\delta)}(x)=H_3(x)-4x \delta,\nonumber\\
H_4^{(\delta)}(x)&=&H_4(x)-32 x^2\delta+12\delta,\quad
H_5^{(\delta)}(x)=H_5(x)-160x^3\delta +200x\delta +48x\delta^2,\nonumber\\
H_6^{(\delta)}(x)&=&H_6(x)-640x^4\delta+x^2(736\delta^2+1680\delta)-
240\delta^2-360\delta,\quad\ldots
\end{eqnarray}
For $B_r$ let us define
\begin{equation}
\bar{q}_j=\sqrt{ag_L}\,y_j,\quad \alpha=g_S/g_L-1,\quad \delta=g/a,
\end{equation}
and introduce a degree $r$ polynomial
\begin{equation}
L^{(\delta,\alpha)}_r(x)=(-1)^r\prod_{j=1}^r(x-y_j^2)/r!,
\end{equation}
which is a deformation of the Laguerre polynomial $L^{(\alpha)}_r(x)$.
For lower $n$,
we have
\begin{eqnarray}
L^{(\delta,\alpha)}_n(x)&=&L^{(\alpha)}_r(x),\quad n=0,1,\quad
L^{(\delta,\alpha)}_2(x)=L^{(\alpha)}_2(x)- \frac{\delta} {2} \left(
-2 + 3x - 3\alpha  + 2x\alpha  - {\alpha }^2 \right), \,
     \\
L^{(\delta,\alpha)}_3(x)&=&L^{(\alpha)}_3(x)-\frac{\delta}{6}
  \left(\vTm\!\! -18 + 45\,x -
13\,x^2 - 33\,\alpha  + 42\,x\,\alpha  -
         6\,x^2\,\alpha  - 18\,{\alpha }^2 + 9\,x\,{\alpha }^2 - 3\,{\alpha }^3
\right.\nonumber\\
&&\left. \qquad\qquad\quad-
         12\,\delta  + 22\,x\,\delta  - 22\,\alpha \,\delta  +
         24\,x\,\alpha \,\delta  - 12\,{\alpha }^2\,\delta  +
         6\,x\,{\alpha }^2\,\delta  - 2\,{\alpha }^3\,\delta  \vTm\right), \\
L^{(\delta,\alpha)}_4(x)&=&L^{(\alpha)}_4(x)-
\frac{\delta}{24} \left(\vTm-144 + 504\, x - 280\, x^2 + 34\,
             x^3 - 300\, \alpha + 582\, x\, \alpha - 190\, x^2\,
\alpha \right.\nonumber\\
&&\qquad\qquad+
12\,
             x^3\, \alpha - 210\, {\alpha}^2 + 210\, x\, {\alpha}^2 - 30\,
             x^2\, {\alpha}^2 - 60\, {\alpha}^3 + 24\,
             x\, {\alpha}^3 - 6\, {\alpha}^4 \nonumber\\
&&\qquad\qquad- 264\, \delta + 760\,
             x\, \delta - 241\, x^2\, \delta - 550\, \alpha\, \delta + 950\,
             x\, \alpha\, \delta - 192\,
             x^2\, \alpha\, \delta \nonumber\\
&&\qquad\quad- 385\, {\alpha}^2\, \delta + 366\,
             x\, {\alpha}^2\, \delta - 36\,
             x^2\, {\alpha}^2\, \delta - 110\, {\alpha}^3\, \delta + 44\,
             x\, {\alpha}^3\, \delta - 11\, {\alpha}^4\, \delta\nonumber\\
&&\qquad\quad -
               144\, {\delta}^2 + 300\,
             x\, {\delta}^2 - 300\, \alpha\, {\delta}^2 + 420\,
             x\, \alpha\, {\delta}^2 - 210\, {\alpha}^2\, {\delta}^2 + 180\,
             x\, {\alpha}^2\, {\delta}^2\nonumber\\
&&\qquad\qquad\ \left. - 60\, {\alpha}^3\, {\delta}^2 + 24\,
             x\, {\alpha}^3\, {\delta}^2 -
               6\, {\alpha}^4\, {\delta}^2 \vTm\right),\quad \ldots
\end{eqnarray}
They are not the so-called ``$q$-{\em deformed\/}" Hermite or
Laguerre polynomials
\cite{And-Ask-Roy}. As in the Calogero systems, the $D_r$  is a special case
$g_S=0$ of the $B_r$ theory described by $L^{(\delta,-1)}(x)$,
which has a zero at $x=0$
for all $r$, see (I.4.20).

  It is remarkable that the spectrum of the small oscillations
at equilibrium is {\em completely independent\/} of
the coupling constant $g$, $a$,
$g_L$ or $g_S$. In other words, the spectrum is the
  {\em topological invariant\/}
of the theory. It is solely determined by the root system. In fact,
the spectrum is
\begin{equation}
2(1+e_j),\quad j=1,\ldots,r,
\label{1exp}
\end{equation}
in which $e_j$ is the $j$-th {\em exponent\/} of the root system $\Delta$.
Explicitly, the spectrum is
\begin{eqnarray}
A:&&\qquad 2\times(1,2,\ldots,r+1),
\label{a1exp}\\
B\, (C)\ \& \ BC:&&\qquad 2\times(2,4,\ldots,2r),\\
D:&&\qquad 2\times(2,4,\ldots,2r-2,r).
\label{d1exp}
\end{eqnarray}
(The lowest frequency of $A$ series is due to the
{\em center of mass\/} motion
which is not described by   the data of the root system.)
The situation is essentially the same as in the Calogero systems,
in which the frequencies of the small oscillations are proportional
  to the above formula (\ref{1exp}) and are independent of the
coupling constant(s) of the rational potential  \cite{cs}.
We strongly believe that this very orderly spectrum is
a good evidence for the integrability
of this type of models, as is the case for the Calogero systems.

\subsection{Quadratic  confining potential case}
\label{Rquad}
Let us first write down the explicit forms of the elementary potential
functions
$v$ and $w$. For the simply-laced root systems $A$ and $D$,
the elementary potential functions are:
\begin{eqnarray}
  v(x)=1-ig/x,\quad w(x)=(a+ix)(b+ix),\quad a,\, b, \, g>0.
\label{ADVform}
\end{eqnarray}
The elementary potential functions are ($a$,\ $b$,\ $g_L$,\ $g_M$, $g_S >0$):
\begin{eqnarray}
B:&& v(x)=1-ig_L/x,\quad w(x)=(a+ix)(b+ix)(1-ig_S/x),\\
C:&&
v(x)=1-ig_S/x,\quad w(x)=(a+2ix)(b+2ix)(1-ig_L/{2x}),\label{Cpot}\\
BC:&&
v(x)=1-ig_M/x,\quad w(x)=(a+ix)(b+ix)(1-ig_L/{2x})(1-ig_S/x).\label{BCpot}
\end{eqnarray}
The $C$ system is slightly different from the one given by van Diejen
\cite{vanDiejen2}, since the latter is the quantum theory.
In the limit $\hbar\to0$, (\ref{Cpot}) is the same as van Diejen's.
Here again those based on $B$ and $C$
systems are equivalent in terms of the overall scaling of the potential
and scaling of the coupling constants. This fact is reflected in their spectra
(\ref{BCalspec}), (\ref{CCalspec}).
In all these cases the `potential' $P$ (\ref{potP})
grows quadratically in $|q|$ as
$|q|\to\infty$.
In the present  case the identities (\ref{Viden}) are replaced by
\begin{equation}
\sum_{j}\left\{V_j(q)+V_j(q)^*\right\}=
\left\{
\begin{array}{rll}
-2(r+1)q^2+const,&& A_r\\[4pt]
-2rq^2+const,&& B_r,\ (C_r),\ D_r,
\end{array}
\right.
\label{VQiden}
\end{equation}
except for the $BC$ case.

Again the equations determining the equilibrium (\ref{equeq}) are {\em Bethe
ansatz\/}-like equations which are expected to determine
two-parameter deformation
of the Hermite and Laguerre polynomials. These will be discussed elsewhere.
  The
spectrum of the small oscillations at equilibrium has a very simple form.
Explicitly, the
spectrum is
\begin{eqnarray}
A:&&\quad 2 j\left[a+b+g\left(r-(j-1)/2\right)\right],\hspace{14mm} j=1,2,
\ldots,r+1,
\label{caspec}\\
B:&&\quad 4 j\left[a+b+g_S+g_L(2r-j-1)\right],\hspace{8.5mm} j=1,2,\ldots,r,
\label{BCalspec}\\
C:&&\quad 8 j\left[a+b+g_L+2g_S(2r-j-1)\right],\quad\ \, j=1,2,\ldots,r,
\label{CCalspec}\\
BC:&&\quad 4 j\left[a+b+g_L/2+g_S+g_M(2r-j-1)\right],\quad\ \, j=1,2,
\ldots,r,\label{BCCalspec}\\
D:&&\quad 4 j\left[a+b+g(2r-j-1)\right],
\hspace{19.2mm} j=1,2,\ldots,r-1,\nonumber\\
&&\quad \mbox{and}\quad 2 r(a+b+g(r-1)).
\label{cdspec}
\end{eqnarray}
The Calogero models are obtained in the singular limit,
$a, b\to\infty$ and by division by $ab$.
In this limit, the above spectrum of small oscillations,
(\ref{caspec})--(\ref{cdspec}) will be proportional to
those of the Calogero models, {\em i.e.\/} (\ref{1exp}),
(\ref{a1exp})--(\ref{d1exp}), as expected.

It is interesting to compare the above spectrum of small oscillations with the
{\em quantum\/} energy eigenvalues. The quantum spectrum
is given by van Diejen \cite{vanDiejen1}:
\begin{eqnarray}
A:&&E_{\vec{n}}=\sum_{1\leq j\leq r+1}n_j\left[\vTmm n_j+2(a+b)-1
+2g(r+1-j)\right],
\label{qcaspec}\\
B:&&E_{\vec{n}}=4\sum_{1\leq j\leq r}n_j\left[\vTmm n_j+a+b-1
+g_S+2g_L(r-j)\right],
\label{qcbspec}\\
C:&&E_{\vec{n}}=8\sum_{1\leq j\leq r}n_j\left[\vTmm n_j+a+b-1
+g_L+4g_S(r-j)\right],
\label{qccspec}\\
BC:&&E_{\vec{n}}=4\sum_{1\leq j\leq r}n_j\left[\vTmm n_j+a+b-1
+g_L/2+g_S+2g_M(r-j)\right],
\label{qcbcspec}
\end{eqnarray}
in which $\vec{n}=(n_1,\ldots,n_r,(n_{r+1}))$ is a set of
`quantum numbers'  parametrising the eigenstates. They are
non-increasing, non-negative integers ($n_1\ge n_2\ge\cdots
n_r\ge (n_{r+1})\ge0$).
In fact, the $B$ and $C$ formulas are special cases of the $BC$ formula.
However, the $D$ formula needs yet to be derived.
The $r, (r+1)$
independent  `lowest lying' modes corresponding to the quantum numbers
\begin{equation}
\vec{n}=(1,0,\ldots,0),(1,1,0,\ldots,0),\ldots,
(1,1,\ldots,1,0),(1,1,\ldots,1),
\end{equation}
have energies,
\begin{eqnarray}
A:&&2j\left[\vTmm a+b+g(r-(j-1)/2)\right],\qquad\qquad\qquad\quad
j=1,\ldots,r+1,
\label{qcaspeclow}\\
B:&&4j\left[\vTmm a+b+g_S+g_L(2r-1-j)\right],\qquad\qquad\quad\;\,
j=1,\ldots,r,
\label{qcbspeclow}\\
C:&&8j\left[\vTmm a+b+g_L+2g_S(2r-1-j)\right],\qquad\qquad\quad j=1,\ldots,r,
\label{qccspeclow}\\
BC:&&4j\left[\vTmm a+b+g_L/2+g_S+g_M(2r-1-j)\right],\qquad j=1,\ldots,r,
\label{qcbcspeclow}
\end{eqnarray}
which are exactly the same as (\ref{caspec})--(\ref{BCCalspec}).

\section{Ruijsenaars-Sutherland systems}
\label{R-Sut}
\setcounter{equation}{0}

The discrete  analogue of the Sutherland systems \cite{Sut}, to be called
Ruijsenaars-Sutherland systems, was introduced originally by
Ruijsenaars and Schneider \cite{Ruij-Sch} for the $A$ type root system.
The quantum eigenfunctions of the $A$ type Ruijsenaars-Sutherland systems
are called Macdonald  polynomials \cite{Macdonald}, which are a one-parameter
deformation of the Jack polynomials \cite{Jack}.
Here we will discuss Ruijsenaars-Sutherland systems for all the classical root
systems,
$A$, $B$, $C$, $D$ and $BC$ \cite{vanDiejen2}.
The structure of the functions $\{V_j(q)\}$, (\ref{AVform}) and
(\ref{otherVform})
are the same as in  the Ruijsenaars-Calogero systems,
but the elementary potential functions $v$ and $w$ are trigonometric instead of
rational.
Because of the identity $\sum_{j=1}^r\left\{V_j(q)+V_j(q)^*\right\}=const.$,
(\ref{Viden}), the
Hamiltonian (\ref{Ham}) can be replaced by a simpler one
\begin{equation}
H'(p,q)=2\sum_{j=1}^r\cosh p_j\,\sqrt{V_j(q)\,{V}_j^*(q)},
\label{Ham3}
\end{equation}
which is obviously positive definite.
This is also valid  for the $BC$ case in contrast to the rational potential
cases discussed in the preceding section \ref{R-Cal}.

The elementary potential functions $v$ and $w$ are:
\begin{eqnarray}
A,\,D:\quad v(x)&=&\cosh\gamma-i\sinh\gamma\cot x,
\qquad w(x)\equiv1,\\
B:\quad v(x)&=&\cosh\gamma_L-i\sinh\gamma_L\cot x,
\quad w(x)=\cosh\gamma_S-i\sinh\gamma_S\cot x,\\
C:\quad v(x)&=&\cosh\gamma_S-i\sinh\gamma_S\cot x,
\quad w(x)=\cosh\gamma_L-i\sinh\gamma_L\cot 2x,\\
C':\quad v(x)&=&\cosh\gamma_S-i\sinh\gamma_S\cot x,
\quad w(x)=(\cosh\gamma_L-i\sinh\gamma_L\cot 2x)^2,\\
BC:\quad v(x)&=&\cosh\gamma_M-i\sinh\gamma_M\cot x,\nonumber\\
&& w(x)=(\cosh\gamma_S-i\sinh\gamma_S\cot x)
(\cosh\gamma_L-i\sinh\gamma_L\cot 2x),
\end{eqnarray}
in which $\gamma_L$, $\gamma_M$ and $\gamma_S$ are
the positive coupling constants for
the long, middle and short roots, respectively.
Both $C$ and $C'$ and $BC$ cases are special cases of the most
general integrable interactions including the long roots
($\alpha_L^2=4$) introduced by van Diejen \cite{vanDiejen0}.
Note that in our paper only the classical dynamics $\hbar\to0$\ is discussed.
That is, van Diejen's  constant $\gamma=i\beta\hbar/2$ (eq.(2.7) in
\cite{vanDiejen0}) is treated as vanishing.

For the systems based on $A$ type root system, the above identity
(\ref{Viden})
\begin{equation}
A_r:\qquad \sum_{j=1}^{r+1}V_j(q)=\sinh[(r+1)\gamma]/\sinh\gamma
\label{ascon}
\end{equation}
is known in a different context \cite{Calbook}.
For the other root systems the identity (\ref{Viden}) reads
\begin{equation}
\sum_{j=1}^{r}\left\{V_j(q)+V_j(q)^*\right\}=
\left\{
\begin{array}{lr}
  2\sinh [r\gamma_L]\cosh[(r-1)\gamma_L+\gamma_S]/\sinh\gamma_L, & B_r,
\\[6pt]
  2\sinh [r\gamma_S]\cosh[(r-1)\gamma_S+\gamma_L]/\sinh\gamma_S, & C_r,
\\[6pt]
  \sinh [(2r-1)\gamma]/\sinh\gamma+1, & D_r.
\label{dscon}
\end{array}
\right.
\end{equation}
It is easy to verify that the $B_r$ formula reduces to the $D_r$ one for
$\gamma_L\to\gamma$ and $\gamma_S=0$.
Had we started from the simplified Hamiltonian
(\ref{Ham3}) instead of the original
one (\ref{Ham}), the above constants (\ref{ascon})--(\ref{dscon})
would give the minimal energies. In the simply-laced $A$ and $D$ cases,
the r.h.s of (\ref{ascon}) and (\ref{dscon}) can be interpreted as
``$q$-{\em deformed integer\/}" version of the dimensionality of $A_r$ and
$D_r$ vector representations, $[r+1]_q$ and $1+[2r-1]_q$.

The equilibrium of the $A$ type theories is achieved at
``{\em equally-spaced\/}"
\begin{equation}
\bar{q}=\pi(0,1,\ldots,r-1,r)/(r+1)+\xi (1,1,\ldots,1),\qquad \xi\in\mathbb{R}:
\mbox{arbitrary},
\label{eqspaced}
\end{equation}
configuration, as in the original Sutherland systems.
In all the other cases, the equilibrium and the frequencies of the small
oscillations are determined  by solving the {\em Bethe ansatz\/}-like equations
(\ref{equeq}) numerically. Certain one-parameter deformation of
the Jacobi polynomials \cite{cs} is expected to describe the equilibrium for
  $B$, $C$, $D$
and
$BC$ systems, which will be discussed elsewhere.
The spectrum of the small oscillations at equilibrium has a very
simple form. Explicitly, the spectrum is
\begin{eqnarray}
A:&&\quad 4\sinh[(r+1-j)\gamma]\sinh[j\gamma]/\sinh\gamma,\hspace{14mm}
  j=1,2,\ldots,r+1,
\label{asspec}\\
B:&&\quad
4\sinh[(2r-1-j)\gamma_L+\gamma_S]\sinh[j\gamma_L]/\sinh\gamma_L,\quad
  j=1,2,\ldots,r-1,\nonumber\\
&&\quad\qquad 2\sinh[(r-1)\gamma_L+\gamma_S]\sinh[r\gamma_L]/\sinh\gamma_L,
\label{bsspec}\\
C:&&\quad
4\sinh[(2r-1-j)\gamma_S+\gamma_L]\sinh[j\gamma_S]/\sinh\gamma_S,\quad
  j=1,2,\ldots,r,
\label{csspec}\\
C':&&\quad
4\sinh[(2r-1-j)\gamma_S+2\gamma_L]\sinh[j\gamma_S]/\sinh\gamma_S,\quad
  j=1,2,\ldots,r,
\label{cpsspec}\\
BC:&&\quad
4\sinh[(2r-1-j)\gamma_M+2\gamma_L+\gamma_S]\sinh[j\gamma_M]/\sinh\gamma_M,\
  j=1,2,\ldots,r,
\label{bcsspec}\\
D:&&\quad 4\sinh[(2r-1-j)\gamma]\sinh[j\gamma]/\sinh\gamma,\hspace{8mm}
  j=1,2,\ldots,r-2,\nonumber\\
&&\quad\qquad 2\sinh[(r-1)\gamma]\sinh[r\gamma]/\sinh\gamma,\qquad\qquad
\mbox{twofold degenerate}.
\label{dsspec}
\end{eqnarray}
The spectrum of the $A$ system (\ref{asspec})%
\footnote{This formula was known to S. Ruijsenaars \cite{Ruij-acIII}.}
  reflects the
symmetry of the Dynkin diagram
$j\leftrightarrow r+1-j$.
The twofold degeneracy of the $D$ spectrum (\ref{dsspec}) also reflects the
symmetry of the $D$ Dynkin diagram.

The  original Sutherland models are obtained in the singular limit
in which all the coupling constant(s) become
infinitesimally small: $0<\gamma, \gamma_L,\gamma_M, \gamma_S\ll1$.
In this limit, the spectrum of small oscillations at equilibrium
  will be  linear in the coupling constant(s).
In these limits, the known spectrum of small oscillations obtained
in I is reproduced.
As for $A_r$, the spectrum (\ref{asspec}) becomes
\begin{equation}
A:\qquad 4(r+1-j)j\gamma, \qquad j=1,\ldots,r+1,
\end{equation}
which is  eq.(5.16) of Corrigan-Sasaki \cite{cs},
to be referred to as (I.5.16).
For $B_r$, the spectrum (\ref{bsspec}) in the limit is
\begin{equation}
B_r:\qquad 4[(2r-1-j)\gamma_L+\gamma_S]j, \quad j=1,\ldots,r-1,\quad
2[(r-1)\gamma_L+\gamma_S]r,\ [2],
\end{equation}
which is (I.5.74).
For $C'_r$, the spectrum (\ref{cpsspec}) in the limit is
\begin{equation}
C':\qquad 4[(2r-1-j)\gamma_S+2\gamma_L]j, \qquad j=1,\ldots,r,
\end{equation}
which is (I.5.81).
For $D_r$, the spectrum (\ref{dsspec}) in the limit is
\begin{equation}
D:\qquad 4(2r-1-j)j\gamma, \quad j=1,\ldots,r-2,\quad
2r(r-1)\gamma, \ [2],
\end{equation}
which is (I.5.60). The limiting
spectra provide non-trivial supporting evidences for the formulas
(\ref{asspec})--(\ref{dsspec}).

It is a challenge to understand the connection between the quantum
spectrum of the Ruijsenaars-Sutherland systems, eq. (5.17) of
\cite{Macdonald} and the above spectrum of small oscillations
(\ref{asspec})--(\ref{dsspec}).
It would be interesting to evaluate the eigenvalues of the
Lax matrices at equilibrium as shown for the C-M systems
\cite{cs}. However, the knowledge of the Lax pairs for the
Ruijsenaars-Schneider systems is still quite limited \cite{RSLax}.

\section*{Acknowledgements}
We thank F.\, Calogero,  S.\, Odake and
S. Ruijsenaars for useful discussion.

\end{document}